\documentclass{article}
\usepackage{spconf,amsmath,graphicx,hyperref}
\usepackage{cite}
\usepackage{amssymb,amsfonts}
\usepackage{algorithmic}
\usepackage{graphicx}
\usepackage{textcomp}
\usepackage{xcolor}
\usepackage{tabularx}
\usepackage{multirow}
\usepackage{booktabs}
\usepackage{makecell}

\title{Rethinking Batch Composition: \\ Promoting feature diversity for Speech Enhancement}
\name{Mingchi Hou$^{1 ,2}$, Ina Kodrasi$^{1}$\thanks{This work was supported by the Swiss National Science Foundation project 200021\_215187 on ``Pathological Speech Enhancement".}}
\address{$^{1}$Idiap Research Institute, Switzerland \\$^2$École Polytechnique Fédérale de Lausanne, Switzerland}
\begin{document}
\ninept

\maketitle

\begin{abstract}
Speech enhancement systems are commonly trained using randomly sampled mini-batches, which may contain limited diversity in intrinsic acoustic characteristics of the clean speech signals. Motivated by previous findings linking clean speech characteristics to enhancement performance, we propose a feature-stratified mini-batch sampling strategy that divides training examples into quantiles according to a clean-speech characteristic and enforces equal representation within each mini-batch. Experiments across two large-scale English and Spanish speech datasets and a smaller dataset containing neurotypical and pathological speech demonstrate improvements over standard random mini-batch sampling across multiple objective enhancement metrics. Ablation studies demonstrate that the gains are robust across training runs, while their magnitude depends on the acoustic characteristic used for stratification. The proposed strategy requires no modification to the enhancement model or training objective, providing a simple and complementary approach to speech enhancement training.

\end{abstract}


\section{Introduction}
\label{sec: introduction}
Speech enhancement (SE) aims to improve the quality and intelligibility of speech degraded by noise, reverberation, and other acoustic distortions.
A broad range of deep learning-based approaches has been developed for this task, including discriminative and generative approaches.
Discriminative approaches learn a mapping from degraded to clean speech through, e.g., spectral mask estimation~\cite{CNNGRU_mask} or spectral coefficient estimation~\cite{Sun_2017}, whereas generative approaches formulate SE as a probabilistic generation problem using diffusion~\cite{richter2023sgmse}, flow matching~\cite{flowse_icassp2025}, or Schr\"{o}dinger Bridge~\cite{jukic2024sb}.
Despite their overall strong performance, these approaches can exhibit substantial performance differences across input conditions, with a range of factors, such as e.g., acoustic degradation (noise type, signal-to-noise ratio (SNR))~\cite{SE_noisy_reverb} or characteristics of the target speaker~\cite{Indiana_WASPAA2021}, contributing to this variability.

This challenge has motivated efforts to improve SE robustness and generalization.
For example, robustness to different acoustic conditions has been addressed by explicitly accounting for noise type and SNR~\cite{fu16_interspeech_snr_aware, ge22_interspeech_snr_aware} or by using mixture-of-experts architectures in which different model components can specialize to different acoustic conditions~\cite{sivaraman20_interspeech_moe}. 
To achieve consistent performance across speakers, speaker-aware and personalized SE approaches have been proposed that condition enhancement on speaker embeddings or that adapt models to individual speakers~\cite{Indiana_WASPAA2021, Tencent_ICASSP2022, MS_Interspeech2024, hou_eusipco_2025}.
Beyond explicitly targeting robustness to acoustic conditions or different speakers, improvements to SE generalization have also been pursued through alternative loss functions~\cite{SE_diff_loss_2020}, training targets~\cite{SQA_ASRU2025, fu2026rethinkingtrainingtargetsarchitectures}, and model architectures~\cite{goswami25_interspeech, chao25b_interspeech}, as well as through strategies for improving the quality and scale of training data~\cite{data_curation_asru2025}.

Despite these advances, recent studies have shown that certain speech samples remain challenging for state-of-the-art SE models, even when the noise type, SNR, and speaker are fixed~\cite{saijo25_interspeech_urgent, Zhang2025Lessons, feature_icassp2026}. This suggests that an important source of enhancement difficulty lies in the intrinsic characteristics of the clean speech signal itself. In our previous work~\cite{feature_icassp2026}, we demonstrated that clean speech characteristics such as formant structure and loudness can be strongly correlated with SE performance, even when the noise type, SNR, and speaker are controlled. 
To the best of our knowledge, this source of variability has never been considered in the design of SE training strategies. Specifically, training data are typically presented to SE models through randomly sampled mini-batches, without explicitly considering the distribution of intrinsic speech characteristics within each batch. Although random sampling provides unbiased coverage of the training set over many iterations, individual mini-batches can still differ substantially in their acoustic composition. For speech data with high intrinsic variability, a randomly sampled batch may consequently cover only a limited region of the acoustic feature space, causing individual gradient updates to be dominated by a restricted range of enhancement difficulty. 

In this work, we propose to explicitly control the distribution of intrinsic clean speech characteristics within each mini-batch to improve SE training and performance. To this end, we propose a simple plug-and-play feature-stratified sampling strategy. 
For each training sample, we extract a user-defined clean speech characteristic (such as e.g., the mean formant frequency or the mean loudness) and partition the training data into quantiles based on its value. Each mini-batch is then constructed with an equal number of samples from each quantile. Importantly, the selected characteristic is used solely to determine the composition of the mini-batch and is not provided as an input to the SE model. 
We evaluate the proposed strategy on discriminative and generative SE approaches across two languages, i.e., English and Spanish, and across two speaker populations, i.e., neuortypical and pathological.
Across these various datasets, we observe consistent improvements in multiple objective SE metrics when the proposed feature-stratified sampling strategy is used for training.

The main contributions of this work are:
\begin{enumerate}
\item We introduce feature-stratified batch sampling for SE, a simple plug-and-play training strategy that explicitly promotes diversity in intrinsic clean speech characteristics within each mini-batch.
\item We demonstrate consistent performance improvements across two SE paradigms, two languages, and two speaker groups, using four objective SE metrics.
\item We show that the improvements are robust to stochastic variations in mini-batch sampling, and that their magnitude depends on the clean speech characteristic used for stratification. In particular, characteristics that are strongly associated with SE performance provide greater benefits, suggesting that stratification is most effective when it promotes diversity along acoustic dimensions related to enhancement difficulty.
\end{enumerate}

\section{Speech Enhancement}

The noisy microphone signal $y(t)$ at time $t$ is given by the sum of the clean speech signal $s(t)$ and the noise signal $n(t)$, i.e.,
\begin{equation}
y(t) = s(t) + n(t).
  \label{equation:eq1}
\end{equation}
In the short-time Fourier transform (STFT) domain,~(\ref{equation:eq1}) is given by
\begin{align}
  Y(f,k) = S(f,k) + N(f,k),
  \label{equation:eq2}
\end{align}
where $Y(f,k)$, $S(f,k)$, and $N(f,k)$ denote the complex-valued STFT coefficients of the noisy, clean speech, and noise signals, respectively, and $f$ and $k$ denote the frequency bin and time frame indices.
The objective of SE is to recover the clean speech signal from its noisy observation. Depending on the modeling paradigm, this objective can be formulated as a direct regression problem or as a generative mapping problem between noisy and clean speech. In this work, we consider representative state-of-the-art models from both paradigms, which are briefly described below. Their implementation and training configurations are detailed in Section~\ref{sec: exp_settings}.

\subsection{Speech enhancement models}
\label{sec: se}

\emph{Complex-valued spectrogram-based regression (CR):} \enspace
\mbox{CR} follows a discriminative regression formulation in which a deep neural network predicts the real and imaginary components of the clean speech STFT from the corresponding noisy STFT~\cite{cr_icassp2023}. The predicted complex-valued spectrum is transformed back to the time domain, and the model is optimized using the mean square error between the reconstructed signal $\hat{s}(t)$ and the clean signal $s(t)$.

\emph{Schr\"{o}dinger Bridge-based model (SB):} \enspace
In contrast to direct regression, \mbox{SB} formulates SE as a generative transport problem learning a stochastic path between the noisy observation and the corresponding clean speech representation~\cite{schrodinger1932theorie, jukic2024sb}. Unlike conventional diffusion-based approaches which progressively perturb clean speech toward a predefined noise distribution~\cite{richter2023sgmse}, SB learns this path directly from the paired noisy and clean representations.
The model is trained with a weighted data-prediction objective defined in both the time and STFT domains. Further details of the SB formulation are provided in~\cite{jukic2024sb}.

\subsection{Training procedure}
\label{sec: sota_training}
State-of-the-art SE models are conventionally trained using mini-batches formed by randomly sampling training examples. 
We consider a training set $\mathcal{D}$ consisting of $N$ pairs of clean-noisy examples, i.e.,
$$ \mathcal{D}=\{(s_i,y_i)\}_{i=1}^{N}, $$
where \(s_i\) denotes the $i$-th clean signal and $y_i$ denotes the corresponding noisy observation. Standard SE training constructs a mini-batch by randomly sampling \(B\) examples from \(\mathcal{D}\).
We use this random sampling strategy as the baseline for training the CR and SB models and compare it against the proposed stratified sampling strategy described in Section~\ref{sec: prop}.

\section{Proposed method}
\label{sec: prop}
Conventional random sampling treats all training examples equally, irrespective of their intrinsic speech characteristics. Our previous analysis showed that intrinsic properties of clean speech are correlated with enhancement gains~\cite{feature_icassp2026}, suggesting that the distribution of such characteristics in the training data may influence how effectively an SE model learns across different types of speech. This motivates us to investigate whether explicitly controlling the distribution of intrinsic speech characteristics during training can improve model generalization.
We therefore propose a stratified mini-batch sampling strategy that balances the representation of a (user-defined) clean speech characteristic across training batches. The approach is independent of the specific characteristic used for stratification and modifies only how training examples are sampled. 

\subsection{Speech characteristics}
\label{sec: acoustic_features}
We extract intrinsic speech characteristics from the clean training utterances using the openSMILE toolkit~\cite{os}. Let $f(s_i)$ denote the value of a selected characteristic for the clean signal $s_i$. The characteristic is computed once for each training utterance and is used solely to determine its sampling group, i.e., it is not provided as an input to the enhancement model. In our experiments, we consider the mean third formant frequency $F_3$~(cf. Section~\ref{sec: exp_corrs}) and the standard deviation of the fundamental frequency $f_0$~(cf. Section~\ref{sec: exp_features}) as speech characteristics for stratification.

\subsection{Stratified mini-batch sampling}
For a given batch size $B$, we propose to choose the number of quantiles $J$ such that $B$ is divisible by $J$. The empirical distribution of the selected characteristic $f(s_i)$ over the training set is divided into $J$ quantiles, denoted by $Q_1$, $\ldots$, $Q_J$. Each training example is assigned to exactly one quantile according to its characteristic value.
Each mini-batch is constructed by randomly sampling $B/J$ utterances from each quantile. Sampling is performed without replacement within an epoch, while the examples are reshuffled between epochs. Thus, each mini-batch contains an equal representation of utterances spanning the range of the selected clean speech characteristic, rather than reflecting a potentially imbalanced distribution.

The quantile boundaries are computed once using the training set and remain fixed throughout training. The same sampling procedure can be applied independently to any SE model, allowing the effect of stratified sampling to be evaluated without modifying the underlying SE architectures, training objectives, or optimization procedures.
The proposed strategy can be implemented using a custom sampler in standard deep learning data loading frameworks.

\section{Experimental settings}
\label{sec: exp_settings}
\subsection{Datasets}

\emph{Clean speech signals.} \enspace
To assess the generalizability of our findings across languages, we use two clean speech datasets, i.e., the English Wall Street Journal (WSJ0) dataset~\cite{garofolo_john_s_csr-i_2007_wsj0} and the Crowdsourced Latin American Spanish (CROWD) dataset~\cite{guevara-rukoz_crowdsourcing_2020}. WSJ0 is a commonly used benchmark for SE~\cite{richter2023sgmse, jukic2024sb}, containing $28.6$ hours of speech from $120$ speakers. We use the standard WSJ0 split, with $101$, $10$, and $8$ speakers assigned to the train, validation, and test sets, respectively. 
Since the CROWD dataset is larger, we select subsets of it with a comparable duration and the same number of speakers as the different WSJ0 splits.

To assess the impact of the proposed batch sampling strategy across neurotypical and pathological speaker groups, we additionally use the PC-GITA dataset~\cite{orozco-arroyave_new_2014_pcgita}. PC-GITA contains $2.8$ hours of speech from $50$ individuals with Parkinson's disease and $50$ neurotypical controls. Each speaker produces $12$ utterances, comprising $10$ sentences, one read text, and one monologue. Given the limited amount of available speech, we use speaker-independent $10$-fold cross-validation for the PC-GITA dataset~\cite{hou_eusipco_2025,feature_icassp2026}. In each fold, $80\%$, $10\%$, and $10\%$ of the speakers are assigned to the training, validation, and test sets, respectively.

For each dataset, all recordings are downsampled to $16$ kHz and the SE models are trained and evaluated independently using only the corresponding training, validation, and test data. 

\emph{Noisy mixtures.} \enspace
Noisy mixtures are constructed using four environmental noise categories from CHiME3~\cite{barker_chime3_2015}, i.e., bus, cafe, pedestrian area, and street. Noise recordings are first downsampled to $16$ kHz before mixture generation. A random segment of a randomly selected noise category is added to a clean speech signal. The mixture SNR is sampled uniformly from $-6$ dB to $14$ dB. This procedure follows the noise-mixing protocol commonly adopted in previous SE studies~\cite{cr_icassp2023,jukic2024sb}.

\subsection{Model architectures and training}
\label{sec: exp_arch}

Following~\cite{cr_icassp2023}, signals are transformed to the STFT domain using a window size of $510$ samples, a hop length of $128$ samples, and a Hann window. The STFT coefficients are then compressed using $\alpha=0.5$ and $\beta=0.33$ as in~\cite{jukic2024sb}.

Both the \mbox{CR} and \mbox{SB} models use the NCSN++M U-Net architecture, a lighter configuration of NCSN++~\cite{song_2021_score,cr_icassp2023}. The \mbox{CR} model removes the noise-conditioning layers, resulting in $22.1$ million trainable parameters, whereas the \mbox{SB} model retains these layers, resulting in $25.2$ million trainable parameters. In addition, \mbox{SB} uses exponential moving average with a decay rate of $0.999$~\cite{jukic2024sb}, a variance-exploding noise schedule with $\sigma_{\min}=0.7$ and $\sigma_{\max}=1.82$, and $50$-step SDE sampling.

Training is performed using the Adam optimizer, a learning rate of $10^{-4}$, and a maximum of $1000$ epochs. Training stops if the validation loss does not decrease for $20$ consecutive epochs. 
For the WSJ0 and CROWD dataset, models are trained on an NVIDIA H100 GPU. For the PC-GITA dataset, models are trained on an RTX 3090 GPU.

\subsection{Evaluation}

SE performance is evaluated using four established objective metrics, i.e., extended short-time objective intelligibility (ESTOI)\cite{estoi}, wideband perceptual evaluation of speech quality (PESQ)\cite{torcoli_pesqc2_2025}, frequency-weighted segmental signal-to-noise ratio (fwSSNR)\cite{fwSSNR}, and scale-invariant signal-to-distortion ratio (SI-SDR)\cite{roux_sdr_2019}. The clean speech signal is used as the reference signal for all metrics. To facilitate comparisons across datasets, we report the improvement in these objective metrics, i.e., $\Delta$ESTOI, $\Delta$PESQ, $\Delta$fwSSNR, and $\Delta$SI-SDR, relative to the noisy mixture.

\begin{table*}[t!]
\centering
\caption{Performance [mean $\pm$ $95\%$ confidence interval] of the considered SE models trained with random batch sampling and the proposed stratified mini-batch sampling. The acoustic feature used for stratification is the mean value of the third formant frequency $F_3$. Models are separately trained and tested on the English WSJ0, Spanish CROWD, and Spanish PC-GITA datasets. For PC-GITA, test performance is reported separately for neurotypical and pathological speakers.}
\label{tbl:WSJ0_CROWD_GITA}
\setlength{\tabcolsep}{1pt}
\begin{tabularx}{\textwidth}{c|c||*{4}{>{\centering\arraybackslash}X}|*{4}{>{\centering\arraybackslash}X}@{}}
\toprule
\multirow{2}{*}{Dataset} & \multirow{2}{*}{Batch} 
    & \multicolumn{4}{c|}{CR model}
    & \multicolumn{4}{c}{SB model} \\
\cmidrule(lr){3-6} \cmidrule(lr){7-10}
& & $\Delta$E-STOI & $\Delta$PESQ & $\Delta$fwSSNR & $\Delta$SI-SDR
  & $\Delta$E-STOI & $\Delta$PESQ & $\Delta$fwSSNR & $\Delta$SI-SDR \\
\midrule
\multirow{2}{*}{WSJ0} 
    & Random & $0.25 \pm 0.01$ & $1.42 \pm 0.02$ 
           & $5.16 \pm 0.10$ & $11.81 \pm 0.21$
           & $0.25 \pm 0.01$ & $1.55 \pm 0.03$
           & $6.13 \pm 0.10$ & $10.59 \pm 0.22$ \\
    & Stratified & $\bf 0.32 \pm 0.01$ & $\bf 1.48 \pm 0.03$ 
           & $\bf 6.17 \pm 0.16$ & $\bf 11.95 \pm 0.21$
           & $\bf 0.26 \pm 0.01$ & $\bf 1.59 \pm 0.03$
           & $\bf 6.32 \pm 0.10$ & $\bf 10.99 \pm 0.22$ \\
\midrule
\multirow{2}{*}{CROWD} 
    & Random & $0.17 \pm 0.01$ & $1.46 \pm 0.03$ 
           & $4.30 \pm 0.08$ & $11.91 \pm 0.18$
           & $0.16 \pm 0.01$ & $1.43 \pm 0.03$
           & $5.42 \pm 0.09$ & $8.71 \pm 0.19$ \\
    & Stratified & $\bf 0.18 \pm 0.01$ & $\bf 1.52 \pm 0.03$ 
           & $\bf 4.69 \pm 0.08$ & $\bf 12.02 \pm 0.18$
           & $\bf 0.18 \pm 0.01$ & $\bf 1.60 \pm 0.03$
           & $\bf 6.33 \pm 0.08$ & $\bf 10.66 \pm 0.18$ \\
           \midrule
\multirow{2}{*}{\makecell{PC-GITA\\Neurotypical}} 
    & Random & $0.16 \pm 0.01$ & $1.25 \pm 0.04$ 
           & $5.73 \pm 0.14$ & $8.65 \pm 0.26$
           & $\bf 0.19 \pm 0.01$ & $1.46 \pm 0.04$
           & $6.47 \pm 0.14$ & $8.52 \pm 0.25$ \\
    & Stratified & $\bf 0.23 \pm 0.01$ & $\bf 1.27 \pm 0.04$ 
           & $\bf 6.30 \pm 0.19$ & $\bf 8.89 \pm 0.24$
           & $\bf 0.19 \pm 0.01$ & $\bf 1.47 \pm 0.04$
           & $\bf 6.49 \pm 0.14$ & $\bf 8.53 \pm 0.25$ \\
           \midrule
\multirow{2}{*}{\makecell{PC-GITA\\Pathological}} 
    & Random & $0.14 \pm 0.01$ & $1.12 \pm 0.04$ 
           & $5.18 \pm 0.18$ & $7.88 \pm 0.27$
           & $0.15 \pm 0.01$ & $1.26 \pm 0.04$
           & $5.71 \pm 0.17$ & $7.53 \pm 0.25$ \\
    & Stratified & $\bf 0.19 \pm 0.01$ & $\bf 1.14 \pm 0.04$ 
           & $\bf 5.62 \pm 0.19$ & $\bf 8.15 \pm 0.23$
           & $\bf 0.15 \pm 0.01$ & $\bf 1.27 \pm 0.04$
           & $\bf 5.75 \pm 0.17$ & $\bf 7.58 \pm 0.25$ \\
\bottomrule
\end{tabularx}
\end{table*}

\section{Experimental Results}
This section evaluates the effectiveness of the proposed stratified mini-batch sampling strategy and examines the factors that influence its performance and robustness. We first compare the proposed strategy with standard random mini-batch sampling across the considered SE models and speech datasets (Section \ref{sec: exp_corrs}). For these main experiments, the mean value of $F_3$ is used as the acoustic characteristic for stratification, motivated by its strong correlation with SE performance observed in our previous work~\cite{feature_icassp2026}. We then conduct two ablation studies. First, we investigate whether the choice of acoustic characteristic affects the benefit of stratification by comparing mean $F_3$ with the standard deviation of the fundamental frequency $f_0$, which showed little or no correlation with SE performance in our previous work (Section \ref{sec: exp_features}). Second, we assess whether the observed improvements are robust to stochastic variations in mini-batch composition by comparing random and stratified sampling across multiple training runs (Section \ref{sec: exp_seeds}).

\subsection{SE performance using standard and stratified mini-batch sampling}
\label{sec: exp_corrs}

\begin{table}[t!]
\centering
\caption{Performance [mean $\pm$ $95\%$ confidence interval] of the CR model trained with stratified mini-batch sampling using the mean $F_3$ and the standard deviation of $f_0$ as stratification features on the English WSJ0 dataset.}
\label{tbl:perf_loud_f0}
\setlength{\tabcolsep}{1pt}
\begin{tabularx}{\columnwidth}{c||*{4}{>{\centering\arraybackslash}X}}
\toprule  
Feature & $\Delta$E-STOI & $\Delta$PESQ & $\Delta$fwSSNR & $\Delta$SI-SDR \\
\midrule
mean $F_3$ 
    & $\bf 0.32 \pm 0.01$ 
    & $\bf 1.48 \pm 0.03$ 
    & $\bf 6.17 \pm 0.16$ 
    & $\bf 11.95 \pm 0.21$ \\
std $f_0$  
    & $0.25 \pm 0.01$ 
    & $\bf 1.48 \pm 0.03$ 
    & $5.17 \pm 0.09$ 
    & $\bf 11.95 \pm 0.21$ \\
\bottomrule
\end{tabularx}
\end{table}

\begin{table}[t!]
\centering
\caption{Performance [mean (std)] of the CR model trained with random batch sampling and the proposed stratified mini-batch sampling with three different seed. The acoustic feature used for stratification is the mean $F_3$ on the English WSJ0 dataset.}
\label{tbl:seed}
\setlength{\tabcolsep}{1pt}
\begin{tabularx}{\columnwidth}{c||*{4}{>{\centering\arraybackslash}X}}
\toprule  
Batch & $\Delta$E-STOI & $\Delta$PESQ & $\Delta$fwSSNR & $\Delta$SI-SDR \\
\midrule
Random
    & 0.24 (0.01)
    & 1.40 (0.02)
    & 5.19 (0.04)
    & 11.70 (0.12) \\
Stratified 
    & \bf 0.27 (0.04)
    & \bf 1.49 (0.02)
    & \bf 5.68 (0.43)
    & \bf 11.93 (0.03) \\
\bottomrule
\end{tabularx}
\end{table}
Table~\ref{tbl:WSJ0_CROWD_GITA} compares the performance of the CR and SB models trained using standard random mini-batch sampling and the proposed stratified mini-batch sampling using a batch size $B=8$ and $J = 4$ quantiles. For mini-batch sampling, training utterances are stratified according to their $F_3$ value. 
Models are separately trained and tested on the English WSJ0, Spanish CROWD, and Spanish PC-GITA datasets. For PC-GITA, test performance is reported separately for neurotypical and pathological speakers.

For the WSJ0 dataset, stratified sampling improves the performance of both the CR and SB models, with the largest gains observed for the CR model,  particularly in E-STOI and fwSSNR.
Improvements for the SB model are more modest across all metrics.
For the CROWD dataset, stratified sampling improves the performance of both the CR and SB models, but the improvements are more pronounced for the SB model, particularly for PESQ, fwSSNR, and SI-SDR.
For the PC-GITA dataset, stratified sampling improves the performance of the CR model for both neurotypical and pathological speakers, with the largest gains observed in E-STOI and fwSSNR. In contrast, the SB model achieves performance comparable to random sampling, with only minor differences between the two strategies. The consistent gains of the CR model across both speaker groups indicate that the proposed sampling strategy also benefits pathological speech.

Overall, stratified mini-batch sampling maintains or improves performance across both SE models, all three datasets, and both neurotypical and pathological speaker populations, with the magnitude of the gains depending on the dataset, model architecture, and evaluation metric. The strategy requires virtually no additional computational cost and leaves the enhancement model, training objective, data, and optimization procedure unchanged, modifying only how training examples are selected to form mini-batches. Therefore, it provides a simple and complementary means of improving SE training and performance without additional model complexity.

\subsection{Effect of the acoustic characteristic used for stratification}
\label{sec: exp_features}

The mean $F_3$ was selected as the stratification feature for the results in Section~\ref{sec: exp_corrs} based on its strong correlation with SE performance reported in our previous work~\cite{feature_icassp2026}. To examine whether the effectiveness of the proposed sampling strategy depends on this correlation, we additionally consider the standard deviation of the fundamental frequency $f_0$ as the stratification feature, which exhibits little or no correlation with SE performance~\cite{feature_icassp2026}. As an exemplary setting for this ablation study, we consider the CR model trained and evaluated on the English WSJ0 dataset.

Table~\ref{tbl:perf_loud_f0} compares the performance of the CR model on WSJ0 when using mean $F_3$ and the standard deviation of $f_0$ for stratification. The mean $F_3$ results are repeated from Section~\ref{sec: exp_corrs} to facilitate a direct comparison.
It can be observed that stratification based on mean $F_3$ yields considerable improvements over the standard deviation of $f_0$ for E-STOI and fwSSNR, whereas PESQ and SI-SDR remain comparable. This suggests that the effectiveness of stratification is not determined simply by introducing greater acoustic diversity within a mini-batch. Instead, the diversity appears to be more beneficial when it is introduced along a dimension that reflects differences in enhancement difficulty. 

\subsection{Robustness to stochastic mini-batch sampling}
\label{sec: exp_seeds}

The results in Section~\ref{sec: exp_corrs} are obtained from a single training run. Since stochastic mini-batch sampling can lead to different mini-batch compositions across training runs, we additionally examine whether the observed performance differences between random and stratified sampling are robust to such variations. To this end, we train the CR model on the English WSJ0 dataset using three independent training runs, with the same random seeds applied to both sampling strategies. For stratified sampling, the mean $F_3$ is used as the stratification feature. 

Table~\ref{tbl:seed} reports the resulting performance across the three runs.
The results show that stratified sampling consistently outperforms random sampling across all four metrics. The relatively small standard deviations across runs indicate that these improvements are consistent across different random seeds, suggesting that the benefit of stratified sampling is robust to stochastic variation in mini-batch composition.

\section{Conclusion and Outlook}

In this paper, we proposed a simple feature-stratified mini-batch sampling strategy for training SE models. The strategy divides the training data according to a clean-speech characteristic strongly correlated with SE performance, i.e., the mean $F_3$, and enforces equal representation of its quantiles within each mini-batch. The proposed strategy improves SE performance across multiple models, datasets, and evaluation metrics, although the magnitude of the gains varies across experimental conditions.
The proposed method requires no modification to the enhancement architecture, loss function, or training procedure, and can be readily incorporated into existing training pipelines. Future work will explore joint diversification based on multiple acoustic characteristics, curriculum learning strategies, effects of different batch sizes and number of quantiles, as well as subjective evaluation of the resulting enhanced speech.

\section{Compliance with Ethical Standards}
This research study was conducted retrospectively using human subject data made available in~\cite{garofolo_john_s_csr-i_2007_wsj0, guevara-rukoz_crowdsourcing_2020, orozco-arroyave_new_2014_pcgita}. Ethical approval was not required as confirmed by the license attached with the data.

\bibliographystyle{IEEEtran}
\bibliography{refs}

\begin{thebibliography}{10}
\providecommand{\url}[1]{#1}
\csname url@samestyle\endcsname
\providecommand{\newblock}{\relax}
\providecommand{\bibinfo}[2]{#2}
\providecommand{\BIBentrySTDinterwordspacing}{\spaceskip=0pt\relax}
\providecommand{\BIBentryALTinterwordstretchfactor}{4}
\providecommand{\BIBentryALTinterwordspacing}{\spaceskip=\fontdimen2\font plus
\BIBentryALTinterwordstretchfactor\fontdimen3\font minus \fontdimen4\font\relax}
\providecommand{\BIBforeignlanguage}[2]{{%
\expandafter\ifx\csname l@#1\endcsname\relax
\typeout{** WARNING: IEEEtran.bst: No hyphenation pattern has been}%
\typeout{** loaded for the language `#1'. Using the pattern for}%
\typeout{** the default language instead.}%
\else
\language=\csname l@#1\endcsname
\fi
#2}}
\providecommand{\BIBdecl}{\relax}
\BIBdecl

\bibitem{CNNGRU_mask}
M.~Hasannezhad, Z.~Ouyang, W.-P. Zhu, and B.~Champagne, ``{An Integrated CNN-GRU Framework for Complex Ratio Mask Estimation in Speech Enhancement},'' in \emph{Proc. Asia-Pac. Signal Inf. Process. Assoc. Annu. Summit Conf.}, {Auckland, New Zealand}, Dec. 2020, pp. 764--768.

\bibitem{Sun_2017}
L.~Sun, J.~Du, L.-R. Dai, and C.-H. Lee, ``Multiple-target deep learning for {LSTM-RNN} based speech enhancement,'' in \emph{Proc. Hands-free Speech Commun. Microphone Arrays}, {San Francisco, USA}, Mar. 2017, pp. 136--140.

\bibitem{richter2023sgmse}
J.~Richter, S.~Welker, J.-M. Lemercier, B.~Lay, and T.~Gerkmann, ``{Speech Enhancement and Dereverberation with Diffusion-Based Generative Models},'' \emph{IEEE/ACM Trans. Audio, Speech, Language Process.}, vol.~31, pp. 2351--2364, Jun. 2023.

\bibitem{flowse_icassp2025}
S.~Lee, S.~Cheong, S.~Han, and J.~W. Shin, ``{FlowSE: Flow Matching-based Speech Enhancement},'' in \emph{Proc. IEEE Int. Conf. Acoust., Speech, Signal Process.}, {Hyderabad, India}, Apr. 2025, pp. 1--5.

\bibitem{jukic2024sb}
A.~Juki{\'c}, R.~Korostik, J.~Balam, and B.~Ginsburg, ``{Schrödinger Bridge for Generative Speech Enhancement},'' in \emph{Proc. Interspeech}, {Kos Island, Greece}, Sep. 2024, pp. 1175--1179.

\bibitem{SE_noisy_reverb}
P.~Gonzalez, T.~S. Alstrøm, and T.~May, ``{Assessing the Generalization Gap of Learning-Based Speech Enhancement Systems in Noisy and Reverberant Environments},'' \emph{IEEE/ACM Trans. Audio, Speech, Language Process.}, vol.~31, pp. 3390--3403, Sep. 2023.

\bibitem{Indiana_WASPAA2021}
S.~Kim and M.~Kim, ``{Test-Time Adaptation Toward Personalized Speech Enhancement: Zero-Shot Learning with Knowledge Distillation},'' in \emph{Proc. IEEE Workshop Appl. Signal Process. Audio Acoust.}, {New Paltz, USA}, Oct. 2021, pp. 176--180.

\bibitem{fu16_interspeech_snr_aware}
S.-W. Fu, Y.~Tsao, and X.~Lu, ``{SNR-Aware Convolutional Neural Network Modeling for Speech Enhancement},'' in \emph{{Proc. Interspeech}}, {San Francisco, USA}, Sep. 2016, pp. 3768--3772.

\bibitem{ge22_interspeech_snr_aware}
X.~Ge, J.~Han, Y.~Long, and H.~Guan, ``{PercepNet+: A Phase and SNR Aware PercepNet for Real-Time Speech Enhancement},'' in \emph{{Proc. Interspeech}}, {Incheon, Korea}, Sep. 2022, pp. 916--920.

\bibitem{sivaraman20_interspeech_moe}
A.~Sivaraman and M.~Kim, ``{Sparse Mixture of Local Experts for Efficient Speech Enhancement},'' in \emph{{Proc. Interspeech}}, {Shanghai, China}, Oct. 2020, pp. 4526--4530.

\bibitem{Tencent_ICASSP2022}
Y.~Ju, W.~Rao, X.~Yan, Y.~Fu, S.~Lv, L.~Cheng, Y.~Wang, L.~Xie, and S.~Shang, ``{TEA-PSE}: Tencent-ethereal-audio-lab personalized speech enhancement system for {ICASSP} 2022 {DNS} challenge,'' in \emph{Proc. IEEE Int. Conf. Acoust., Speech, Signal Process.}, {Singapore, Singapore}, May 2022, pp. 9291--9295.

\bibitem{MS_Interspeech2024}
T.~Pärnamaa and A.~Saabas, ``Personalized {Speech} {Enhancement} {Without} a {Separate} {Speaker} {Embedding} {Model},'' in \emph{Proc. Interspeech}, {Kos Island, Greece}, Sep. 2024, pp. 4863--4867.

\bibitem{hou_eusipco_2025}
M.~Hou and I.~Kodrasi, ``{Variational Autoencoder for Personalized Pathological Speech Enhancement},'' in \emph{Proc. Eur. Signal Process. Conf.}, {Palermo, Italy}, Sep. 2025, pp. 116--120.

\bibitem{SE_diff_loss_2020}
S.~Braun and I.~Tashev, ``A consolidated view of loss functions for supervised deep learning-based speech enhancement,'' in \emph{Proc. Int. Conf. Telecommun. Signal Process.}, {Brno, Czech}, Jul. 2020.

\bibitem{SQA_ASRU2025}
W.~Wang, W.~Zhang, C.~Li, J.~Shi, S.~Watanabe, and Y.~Qian, ``{Improving Speech Enhancement with Multi-Metric Supervision from Learned Quality Assessment},'' in \emph{Proc. IEEE Autom. Speech Recognit. Understanding}, {Honolulu, USA}, Dec. 2025, pp. 1--8.

\bibitem{fu2026rethinkingtrainingtargetsarchitectures}
\BIBentryALTinterwordspacing
S.-W. Fu, R.~Chao, X.~Yang, S.-F. Huang, R.~E. Zezario, R.~Nasretdinov, A.~Jukić, Y.~Tsao, and Y.-C.~F. Wang, ``Rethinking training targets, architectures and data quality for universal speech enhancement,'' 2026. [Online]. Available: \url{https://arxiv.org/abs/2603.02641}
\BIBentrySTDinterwordspacing

\bibitem{goswami25_interspeech}
N.~Goswami and T.~Harada, ``{FUSE: Universal Speech Enhancement using Multi‐Stage Fusion of Sparse Compression and Token Generation Models for the URGENT 2025 Challenge},'' in \emph{{Proc. Interspeech}}, {Rotterdam, The Netherlands}, Aug. 2025, pp. 883--887.

\bibitem{chao25b_interspeech}
R.~Chao, R.~Nasretdinov, Y.-C.~F. Wang, A.~Jukic, S.-W. Fu, and Y.~Tsao, ``{Universal Speech Enhancement with Regression and Generative Mamba},'' in \emph{{Proc. Interspeech}}, {Rotterdam, The Netherlands}, Aug. 2025, pp. 888--892.

\bibitem{data_curation_asru2025}
C.~Li, W.~Zhang, W.~Wang, R.~Scheibler, K.~Saijo, S.~Cornell, Y.~Fu, M.~Sach, Z.~Ni, A.~Kumar, T.~Fingscheidt, S.~Watanabe, and Y.~Qian, ``Less is more: Data curation matters in scaling speech enhancement,'' in \emph{Proc. IEEE Autom. Speech Recognit. Understanding}, {Honolulu, USA}, Dec. 2025, pp. 1--8.

\bibitem{saijo25_interspeech_urgent}
K.~Saijo, W.~Zhang, S.~Cornell, R.~Scheibler, C.~Li, Z.~Ni, A.~Kumar, M.~Sach, Y.~Fu, W.~Wang, T.~Fingscheidt, and S.~Watanabe, ``{Interspeech 2025 URGENT Speech Enhancement Challenge},'' in \emph{{Proc. Interspeech}}, {Rotterdam, The Netherlands}, Aug. 2025, pp. 858--862.

\bibitem{Zhang2025Lessons}
W.~Zhang, K.~Saijo, S.~Cornell, R.~Scheibler, C.~Li, Z.~Ni, A.~Kumar, M.~Sach, W.~Wang, Y.~Fu, S.~Watanabe, T.~Fingscheidt, and Y.~Qian, ``Lessons learned from the {URGENT} 2024 speech enhancement challenge,'' in \emph{Proc. Interspeech}, {Rotterdam, The Netherlands}, Aug. 2025, pp. 853--857.

\bibitem{feature_icassp2026}
M.~Hou and I.~Kodrasi, ``Influence of clean speech characteristics on speech enhancement performance,'' in \emph{Proc. IEEE Int. Conf. Acoust., Speech, Signal Process.}, {Barcelona, Spain}, May 2026, pp. 17\,427--17\,431.

\bibitem{cr_icassp2023}
J.-M. Lemercier, J.~Richter, S.~Welker, and T.~Gerkmann, ``Analysing diffusion-based generative approaches versus discriminative approaches for speech restoration,'' in \emph{Proc. IEEE Int. Conf. Acoust., Speech, Signal Process.}, {Rhodes Island, Greece}, May 2023, pp. 1--5.

\bibitem{schrodinger1932theorie}
E.~Schr\"{o}dinger, ``Sur la th{\'e}orie relativiste de l'{\'e}lectron et l'interpr{\'e}tation de la m{\'e}canique quantique,'' in \emph{Annales de l'institut Henri Poincar{\'e}}, vol.~2, no.~4, 1932, pp. 269--310.

\bibitem{os}
F.~Eyben, M.~W{\"o}llmer, and B.~Schuller, ``{openSMILE} -- {The Munich} versatile and fast open-source audio feature extractor,'' in \emph{Proc. ACM Int. Conf. Multimedia}, {Florence, Italy}, Oct. 2010, pp. 1459--1462.

\bibitem{garofolo_john_s_csr-i_2007_wsj0}
J.~S. Garofolo, D.~Graff, D.~Paul, and D.~Pallett, ``{CSR-I (WSJ0) Complete LDC93S6A},'' Web Download, {Philadelphia: Linguistic Data Consortium}, 1993.

\bibitem{guevara-rukoz_crowdsourcing_2020}
A.~Guevara-Rukoz~et al., ``Crowdsourcing {Latin} {American} {Spanish} for {Low}-{Resource} {Text}-to-{Speech},'' in \emph{Proc. Int. Conf. Lang. Resour. Eval}, {Marseille, France}, May 2020, pp. 6504--6513.

\bibitem{orozco-arroyave_new_2014_pcgita}
J.~R. Orozco-Arroyave, J.~D. Arias-Londoño, J.~F. Vargas-Bonilla, M.~C. González-Rátiva, and E.~Nöth, ``New {Spanish} speech corpus database for the analysis of people suffering from {Parkinson}'s disease,'' in \emph{Proc. Int. Conf. Lang. Resour. Eval}, {Reykjavik, Iceland}, May 2014, pp. 342--347.

\bibitem{barker_chime3_2015}
J.~Barker, R.~Marxer, E.~Vincent, and S.~Watanabe, ``The third ‘{CHiME}’ speech separation and recognition challenge: {Dataset}, task and baselines,'' in \emph{Proc. IEEE Autom. Speech Recognit. Understanding}, {Scottsdale, USA}, Dec. 2015, pp. 504--511.

\bibitem{song_2021_score}
Y.~Song, J.~Sohl-Dickstein, D.~P. Kingma, A.~Kumar, S.~Ermon, and B.~Poole, ``Score-based generative modeling through stochastic differential equations,'' in \emph{Proc. Int. Conf. Learn. Represent.}, {Virtual Event, Austria}, May 2021.

\bibitem{estoi}
J.~Jensen and C.~H. Taal, ``An algorithm for predicting the intelligibility of speech masked by modulated noise maskers,'' \emph{IEEE/ACM Trans. Audio, Speech, Language Process.}, vol.~24, no.~11, pp. 2009--2022, Nov. 2016.

\bibitem{torcoli_pesqc2_2025}
M.~Torcoli, M.~M. Halimeh, and E.~A.~P. Habets, ``{Navigating PESQ: Up-to-Date Versions and Open Implementations},'' in \emph{Proc. Speech Communication; 16th ITG Conference}, {Berlin, Germany}, 2025, pp. 51--55.

\bibitem{fwSSNR}
Y.~Hu and P.~C. Loizou, ``Evaluation of objective quality measures for speech enhancement,'' \emph{IEEE Trans. Audio, Speech, Language Process.}, vol.~16, no.~1, pp. 229--238, Jan. 2008.

\bibitem{roux_sdr_2019}
J.~L. Roux, S.~Wisdom, H.~Erdogan, and J.~R. Hershey, ``{SDR} – {Half}-baked or {Well} {Done}?'' in \emph{Proc. IEEE Int. Conf. Acoust., Speech, Signal Process.}, {Brighton, UK}, May 2019, pp. 626--630.

\end{thebibliography}

\end{document}